\documentclass{llncs}

\usepackage{mathrsfs}
\usepackage[linesnumbered,ruled,vlined]{algorithm2e}
\usepackage{tikz}
\usetikzlibrary{arrows,snakes,backgrounds}

\newcommand{\eq}{\leftarrow}

\newcommand{\argmax}{\mathop{\mathrm{argmax}}}     

\begin{document}
\title{Equilibria on a Game with Discrete Variables}
\titlerunning{Equilibria with Discrete Variables}  
%
\author{Jo\~ao Pedro PEDROSO\inst{1} \and Yves SMEERS\inst{2}}
\authorrunning{Pedroso and Smeers}   
%
%
\institute{  INESC - Porto and\\
  DCC - Faculdade de Ci\^encias, Universidade do Porto, Portugal\\
  \email{jpp@fc.up.pt}
\and
  {CORE, Universit\'e Catholique de Louvain}\\
  \email{Yves.Smeers@uclouvain.be}\\
}
\date{January 2010}

\maketitle              

\begin{abstract}
  Equilibrium in Economics has been seldom addressed in a situation where some variables are discrete.  This work introduces a problem related to lot-sizing with several players, and analyses some strategies which are likely to be found in real world games.  An illustration with a simple example is presented, with concerns about the difficulty of the problem and computation possibilities.
\end{abstract}

\section{Introduction}
\label{sec:introd}
Market equilibrium is a classical problem arising in Economics, with
applications ranging from the analysis of market power to the
simulation of new types of regulation.  The most studied versions
involve agents that choose a (continuous) variable; a Nash equilibrium
occurs when no firm can do better by unilaterally changing its
strategy.  When the variable being played is the quantity, this is
usually called a Cournot equilibrium when all firms play
simultaneously; when a leader firm moves first, and follower firms
move afterwords, this is called a Stackelberg equilibrium (see
e.g.~\cite{Osborne1994} for an introduction).

Much less attention has been drawn to those games where some variables
in this competition are discrete.  In this case, the computation of
equilibria is a much harder task; actually, determining the optimal
strategy for a single player is itself an NP-complete problem in many
situations.

Let us start with the description of the market we will deal with in
this paper.  The demand is forecast with complete certainty for a
number of coming periods.  A firm has the possibility of producing in
a given period or not; demand is met in each period with goods either
produced or existing in inventory.  There are fixed (setup) and
inventory costs, and the production capacity is limited.  The
variables for each firm are, thus:
\begin{itemize}
\item decide whether to produce or not, for each period in the
  planning horizon;
\item choose the quantity to place in the market in each period.
\end{itemize}
If the demand is independent of the price (i.e. it is a fixed amount)
and there is only one firm, this leads to the well known lot-sizing
problem~\cite{Bitran1982}.  We are more interested in the case where demand is
dependent of the price; this dependence will be captured by modelling
demand as a linear function of the price (or, equivalently, price as a
linear function of the quantity put in the market).  This leads to a
lot-sizing variant where the firm, instead of simply meeting the
demand, must decide what quantity to put in the market.  Price
is, thus, a function of the total quantity, i.e., the quantity played
summed for all the firms.

\section{Some interesting markets}
\subsection{Example}
\label{sec:data}

For the sake of clarity, we will use an example throughout this paper.

Demand is different from period to period, and is modelled by:
\begin{equation}
  \label{eq:demand}
  P_t(Q_t) = \max(a_t - b_t Q_t, 0), \;\;\; \textrm{for } t=1,\ldots,T,  
\end{equation}
where $t$ is the period, $T$ is the total number of periods, $a_t$,
$b_t$ are parameters of the model, and $Q_t \geq 0$ is the total
quantity placed in the market in period~$t$.

The decision variables concern producing or not in each period, as
well as the amount to produce, and the amount to place in the market.
Let $y$ be the vector of setup, binary variables, where $y_{t}$ is 1
if there is production during period $t$, and $0$ otherwise.
Variables $x_{t}$ and $q_t$ are, respectively, the corresponding
manufactured amount and the quantity placed in the market in $t$, and
the quantity held in inventory at the end of the period $t$ is $h_t$;
these are non-negative, continuous variables.  The bill of materials
can be written as:
\begin{equation}
  \label{eq:bom}
h_{t-1} + x_t = q_t + h_t, \;\;\; \textrm{for } t=1,\ldots,T.  
\end{equation}
We assume that there is a limit $K$ on the capacity available on each
period, and production can only occur if machines have been setup;
this implies that
\begin{equation}
  \label{eq:setup}
    x_{t} \leq K y_{t}, \;\;\; \textrm{for } t=1,\ldots,T.  
\end{equation}

Let us denote the fixed production costs by $F$ and the unit inventory
costs by $H$.  For a given production plan the total costs are:
$$C(y,h) = \sum_{t=1}^{T} (F y_t + H h_t).$$

We will provide a numerical example, allowing to draw conclusions for
simple situations.  Demand is enough for at least one firm to be able
to produce with profit on the first half of the horizon, and larger
for the second half.  Inventory costs are such that it is worthy to
produce some periods in advance (when demand justifies it), but too
early production is discouraged by them.

We assume throughout this paper that firms know each others'
costs and capacities (i.e., technology is
known).  

\subsection{Monopoly}
\label{sec:monopoly}

In the monopoly case, all the variables are under the control of the
firm.  In this case, market quantities are those decided by the
monopolist, $Q_t = q^M_t$, and the corresponding price is given by
Equation~\ref{eq:demand}.

The profit is therefore given by
\begin{equation}
  \label{eq:monopoly}
  \Pi = \sum_{t=1}^{T} \left[q^M_t P_t(q^M_t) - (F y_t + H h_t)\right].
\end{equation}

For a single period, the optimal result for this model is well known;
maximum profit is obtained when the derivatives with respect to the
quantity are zero, which for linear demand leads to optimal quantities
$$q^{*M} = \frac{a}{2 b}.$$  
The firm will produce if the corresponding profit is positive, i.e.,
if the revenue is larger than the fixed cost~$F$.

When there are several periods, the situation is more complex; an
illustration is provided for our example in Table~\ref{tab:monopoly}.
The firm produces a small quantity for the initial periods, and
produces at full capacity after the demand raises.  Notice that this
problem is NP-hard even when the quantities are fixed~\cite{garey79}
(this is the ``standard'' lot-sizing problem, i.e., all that the firm
has to do is to meet demand at minimal cost), so for more realistic
examples the computation of the optimum is not trivial.  For obtaining
the results presented in this paper we used the software
Couenne~\cite{BelottiCouMan09}, which is based on the latest
developments in mixed-integer nonlinear
programming~\cite{Hemmecke2010}.
\begin{table}[htbp]
  \centering
  \begin{scriptsize}
  \begin{tabular}{|c|c|c||c|r|r|r||c|r|r|r|}
    \hline
    \multicolumn{3}{|c||}{~}
    & \multicolumn{4}{|c||}{$F=10, H=1, K=10$} 
    & \multicolumn{4}{|c|}{$F=10, H=1, K=25$}\\\hline
Period (t) & ~$a_t$~ & ~$b_t$~
& ~~$y_t$~~ & ~~$x_t$~~ & ~~$h_t$~~ & ~~$q_t$~~
& ~~$y_t$~~ & ~~$x_t$~~ & ~~$h_t$~~ & ~~$q_t$~~ 
\\\hline
  1 &    10 &     1 &   1 &  5.00 &  0.00 &  5.00    &   1 & 13.49 &  8.49 &  4.99\\
  2 &    10 &     1 &   1 &  9.50 &  4.50 &  5.00    &   0 &  0.00 &  4.00 &  4.50\\
  3 &    10 &     1 &   0 &  0.00 &  0.00 &  4.50    &   0 &  0.00 &  0.00 &  4.00\\
  4 &    10 &   0.5 &   1 & 10.00 &  0.00 & 10.00    &   1 & 10.00 &  0.00 & 10.00\\
  5 &    10 &   0.5 &   1 & 10.00 &  0.00 & 10.00    &   1 & 19.00 &  9.00 & 10.00\\
  6 &    10 &   0.5 &   1 & 10.00 &  0.00 & 10.00    &   0 &  0.00 &  0.00 &  9.00\\
         \hline
    \multicolumn{3}{|c||}{~}
    & \multicolumn{4}{|c|}{$\Pi = 170.25$}
    & \multicolumn{4}{|c||}{$\Pi = 171.75$} 
\\\hline\hline
  1 &    10 &  0.25 &   1 & 10.00 &  0.00 & 10.00    &   1 & 20.00 &  0.00 & 20.00\\
  2 &    10 &  0.25 &   1 & 10.00 &  0.00 & 10.00    &   1 & 23.01 &  3.01 & 20.00\\
  3 &    10 &  0.25 &   1 & 10.00 &  2.00 &  8.00    &   1 & 25.00 & 10.01 & 18.00\\
  4 &    10 & 0.125 &   1 & 10.00 &  0.00 & 12.00    &   1 & 25.00 &  3.00 & 32.01\\
  5 &    10 & 0.125 &   1 & 10.00 &  0.00 & 10.00    &   1 & 25.00 &  0.00 & 28.00\\
  6 &    10 & 0.125 &   1 & 10.00 &  0.00 & 10.00    &   1 & 25.00 &  0.00 & 25.00\\
         \hline
    \multicolumn{3}{|c||}{~}
    & \multicolumn{4}{|c|}{$\Pi = 429.00$} 
    & \multicolumn{4}{|c||}{$\Pi = 768.875$} 
\\\hline
  \end{tabular}    
  \end{scriptsize}
  \caption{Optimal results for the monopoly situation, under four
    parameter sets.}
  \label{tab:monopoly}
\end{table}

\subsection{Oligopoly}
\label{sec:olipoly}

Let us now consider the case of several firms operating in the market,
an oligopoly.  The difference between the monopoly and the current
situation is that now the total quantity put in the market is no
longer the decision of a single firm.  This can be written in the
profit for each firm $i$ as:
\begin{equation}
  \label{eq:oligopoly}
  \Pi^i = \sum_{t=1}^{T} \left[q^i_t P_t(\sum_{j=1}^{N}q^j_t) - (F^i y^i_t + H^i h^i_t)\right].
\end{equation}
where $N$ is the number of firms operating.

\subsubsection{One-period games.}

For a single period, an equilibrium for this model can again be
derived analytically, and is well known: it is the Cournot
equilibrium.  Assuming that the fixed costs are null or low enough,
maximum profit is obtained when the partial derivatives of the profit
with respect to the quantity are zero, for each firm, leading to the
system of equations
$$q^i = \frac{a + b \sum_{j\neq i}q^j}{2 b}.$$	
At this point, no firm has incentive to deviate; any unilateral
variation will lead to a smaller profit.

The solution is more complex in the presence of fixed costs.  The
previous equilibrium may be a solution with positive profit for all
firms in this case too.  When not all firms can produce with profit,
i.e., fixed costs~$F^i$ are larger than the revenue at the Cournot
equilibrium, there may be the case that one firm can produce with
profit, but if any other enters the market, all will have losses.

Another potential equilibrium, seldom analysed, occurs when one firm
plays a large quantity, in order to try to put the others out of the
market; this ``aggressive'' firm plays a large quantity, in such a way
that the other companies' profit is zero (or negative, if they produce
a positive amount).  Clearly, if all firms play this large amount,
they all will be worse off; but if one of them succeeds imposing this
quantity, as in a Stackelberg equilibrium, the others' optimal
strategy is no production.

Yet another possibility in this game occurs when the firms coalesce
and maximise the sum of the profits of all firms, deciding in another
stance how to share them.  This situation is quite similar to the
monopoly case.

The multi-period situation is considerably more complex.  We present
next some results for equilibria with a single move for all periods;
an \emph{iterated} version will be developed in section~\ref{sec:iterated}.

\subsubsection{Multi-period duopoly: a case study.}

For illustrating the duopoly case we take the same production and
demand parameters used for the monopoly example, and analyse the
behaviour of a market with two firms.  

For the following results we use fixed-point iteration, and the
software Couenne for optimisation.  In each iteration, Firm 1
optimises its quantities based on the previous output of Firm 2, and
vice-versa; details are available in Algorithm~\ref{alg:FPI}, where
$\bar{q}^i$ are initial values for the quantities played,
$\bar{q}^{*i}$ are the corresponding values optimised for the
competitor's current value, and $q^i$ the decision variables in each
optimization problem; $\epsilon$~is a convergence criterion.  Initial
quantities for Firm 2 are zero for all periods, except if otherwise
stated.
\begin{algorithm}
  initialise $\bar{q}^1$ and $\bar{q}^2$\;
  \SetKwRepeat{Repeat}{repeat}{until}
  \Repeat{$\Delta < \epsilon$} {
    $\bar{q}^{*1} \eq \argmax(\Pi^1(q^1, \bar{q}^2))$\;
    $\bar{q}^{*2} \eq \argmax(\Pi^2(\bar{q}^{*1}, q^2))$\;
    $\Delta \eq |\bar{q}^{*1} - \bar{q}^1| +  |\bar{q}^{*2} - \bar{q}^2|$\;
    $\bar{q}^1 \eq \bar{q}^{*1}$\;
    $\bar{q}^2 \eq \bar{q}^{*2}$\;
  }
  \caption{A fix-point iteration for the duopoly equilibrium\label{alg:FPI}}
\end{algorithm}

Let us first analyse the case where firms are symmetric, as in the
results presented in Table~\ref{tab:duopoly}.  The first observation
is that when the setup decisions are important, even though the
profits at equilibrium are roughly the same, the quantities played in
each period by each of the firms may be considerably different.

\begin{table}[htbp]
  \centering
  \begin{scriptsize}
  \begin{tabular}{|c|c|c||c|r|r|r||c|r|r|r|}
    \hline
    \multicolumn{3}{|c||}{~}
    & \multicolumn{4}{|c||}{Firm 1} 
    & \multicolumn{4}{|c|}{Firm 2}\\\hline
    \hline
    \multicolumn{3}{|c||}{~}
    & \multicolumn{4}{|c||}{$F=10, H=1, K=10$} 
    & \multicolumn{4}{|c|}{$F=10, H=1, K=10$}\\\hline
Period (t) & ~$a_t$~ & ~$b_t$~
& ~~$y_t$~~ & ~~$x_t$~~ & ~~$h_t$~~ & ~~$q_t$~~
& ~~$y_t$~~ & ~~$x_t$~~ & ~~$h_t$~~ & ~~$q_t$~~ 
\\\hline  
  1 &  10 &   1&  1 &  8.37 &  5.04 &  3.33	&  1 &  6.33 &  3.00 &  3.34	\\
  2 &  10 &   1&  0 &  0.00 &  2.04 &  3.00	&  0 &  0.00 &  0.00 &  3.00	\\
  3 &  10 &   1&  0 &  0.00 &  0.00 &  2.04	&  1 & 10.00 &  6.08 &  3.92	\\
  4 &  10 & 0.5&  1 & 10.00 &  4.41 &  5.59	&  0 &  0.00 &  0.00 &  6.08	\\
  5 &  10 & 0.5&  0 &  0.00 &  0.00 &  4.41	&  1 & 10.00 &  3.54 &  6.46	\\
  6 &  10 & 0.5&  1 &  8.22 &  0.00 &  8.22	&  0 &  0.00 &  0.00 &  3.54	\\
         \hline
    \multicolumn{3}{|c||}{~}
    & \multicolumn{4}{|c|}{$\Pi^1 = 67.13$} 
    & \multicolumn{4}{|c||}{$\Pi^2 = 65.72$} 
\\\hline\hline
    \multicolumn{3}{|c||}{~}
    & \multicolumn{4}{|c||}{$F=10, H=1, K=25$} 
    & \multicolumn{4}{|c|}{$F=10, H=1, K=25$}\\\hline
Period (t) & ~$a_t$~ & ~$b_t$~
& ~~$y_t$~~ & ~~$x_t$~~ & ~~$h_t$~~ & ~~$q_t$~~
& ~~$y_t$~~ & ~~$x_t$~~ & ~~$h_t$~~ & ~~$q_t$~~ 
\\\hline  
  1 &  10 &   1&  1 &  8.33 &  5.00 &  3.34	&  1 &  6.33 &  3.00 &  3.34	\\
  2 &  10 &   1&  0 &  0.00 &  2.00 &  3.00	&  0 &  0.00 &  0.00 &  3.00	\\
  3 &  10 &   1&  0 &  0.00 &  0.00 &  2.00	&  1 &  9.33 &  5.33 &  4.00	\\
  4 &  10 & 0.5&  1 & 12.67 &  5.33 &  7.34	&  0 &  0.00 &  0.00 &  5.33	\\
  5 &  10 & 0.5&  0 &  0.00 &  0.00 &  5.33	&  1 & 12.66 &  5.33 &  7.33	\\
  6 &  10 & 0.5&  1 &  7.34 &  0.00 &  7.34	&  0 &  0.00 &  0.00 &  5.33	\\
         \hline
    \multicolumn{3}{|c||}{~}
    & \multicolumn{4}{|c|}{$\Pi^1 = 62.15$} 
    & \multicolumn{4}{|c||}{$\Pi^2 = 61.43$} 
\\\hline\hline
    \multicolumn{3}{|c||}{~}
    & \multicolumn{4}{|c||}{$F=10, H=1, K=10$} 
    & \multicolumn{4}{|c|}{$F=10, H=1, K=10$}\\\hline
Period (t) & ~$a_t$~ & ~$b_t$~
& ~~$y_t$~~ & ~~$x_t$~~ & ~~$h_t$~~ & ~~$q_t$~~
& ~~$y_t$~~ & ~~$x_t$~~ & ~~$h_t$~~ & ~~$q_t$~~ 
\\\hline  
  1 &  10 & 0.25&  1 & 10.00 &  0.00 & 10.00	&  1 & 10.00 &  0.00 & 10.00	\\
  2 &  10 & 0.25&  1 & 10.00 &  0.89 &  9.11	&  1 & 10.00 &  0.90 &  9.10	\\
  3 &  10 & 0.25&  1 & 10.00 &  3.11 &  7.78	&  1 & 10.00 &  3.11 &  7.78	\\
  4 &  10 & 0.125&  1 & 10.00 &  0.22 & 12.89	&  1 & 10.00 &  0.23 & 12.89	\\
  5 &  10 & 0.125&  1 & 10.00 &  0.00 & 10.22	&  1 & 10.00 &  0.00 & 10.23	\\
  6 &  10 & 0.125&  1 & 10.00 &  0.00 & 10.00	&  1 & 10.00 &  0.00 & 10.00	\\
         \hline
    \multicolumn{3}{|c||}{~}
    & \multicolumn{4}{|c|}{$\Pi^1 = 321.375$} 
    & \multicolumn{4}{|c||}{$\Pi^2 = 321.368$} 
\\\hline\hline
    \multicolumn{3}{|c||}{~}
    & \multicolumn{4}{|c||}{$F=10, H=1, K=25$} 
    & \multicolumn{4}{|c|}{$F=10, H=1, K=25$}\\\hline
Period (t) & ~$a_t$~ & ~$b_t$~
& ~~$y_t$~~ & ~~$x_t$~~ & ~~$h_t$~~ & ~~$q_t$~~
& ~~$y_t$~~ & ~~$x_t$~~ & ~~$h_t$~~ & ~~$q_t$~~ 
\\\hline  
  1 &  10 & 0.25&  1 & 13.34 &  0.00 & 13.34	&  1 & 13.34 &  0.00 & 13.34	\\
  2 &  10 & 0.25&  1 & 13.34 &  0.00 & 13.34	&  1 & 13.34 &  0.00 & 13.34	\\
  3 &  10 & 0.25&  1 & 13.34 &  0.00 & 13.34	&  1 & 13.34 &  0.00 & 13.34	\\
  4 &  10 & 0.125&  1 & 25.00 &  0.00 & 25.00	&  1 & 25.00 &  0.00 & 25.00	\\
  5 &  10 & 0.125&  1 & 25.00 &  0.00 & 25.00	&  1 & 25.00 &  0.00 & 25.00	\\
  6 &  10 & 0.125&  1 & 25.00 &  0.00 & 25.00	&  1 & 25.00 &  0.00 & 25.00	\\
         \hline
    \multicolumn{3}{|c||}{~}
    & \multicolumn{4}{|c|}{$\Pi^1 = 354.558$} 
    & \multicolumn{4}{|c||}{$\Pi^2 = 354.55$} 
\\\hline
  \end{tabular}
\end{scriptsize}
\caption{Results for a duopoly: Cournot equilibria for the
    lot-sizing problem.}
  \label{tab:duopoly}
\end{table}

Notice that for small demand, when capacities increase, both firms may
become worse off even if the costs are unchanged; this has occurred
from the topmost to the second situations in Table~\ref{tab:duopoly}.
When demand is large enough, this no longer occurs (third and fourth
entries in the table).

It is common to have several equilibria on games with discrete
variables; an illustration with our example is presented in
Table~\ref{tab:manyeq}.  This tables shows two different outcomes of
the fixed-point iteration, obtained using different starting
solutions.  Both firms are better off in the top scenario, even though
no one has incentive to deviate from the situation in the bottom.

\begin{table}[htbp]
  \centering
  \begin{scriptsize}
  \begin{tabular}{|c|c|c||c|r|r|r||c|r|r|r|}
    \hline
    \multicolumn{3}{|c||}{~}
    & \multicolumn{4}{|c||}{Firm 1} 
    & \multicolumn{4}{|c|}{Firm 2}\\\hline
    \hline
    \multicolumn{3}{|c||}{~}
    & \multicolumn{4}{|c||}{$F=10, H=1, K=10$} 
    & \multicolumn{4}{|c|}{$F=10, H=1, K=10$}\\\hline
Period (t) & ~$a_t$~ & ~$b_t$~
& ~~$y_t$~~ & ~~$x_t$~~ & ~~$h_t$~~ & ~~$q_t$~~
& ~~$y_t$~~ & ~~$x_t$~~ & ~~$h_t$~~ & ~~$q_t$~~ 
\\\hline  
  1 &  10 &   1&  1 &  8.37 &  5.04 &  3.33	&  1 &  6.33 &  3.00 &  3.34	\\
  2 &  10 &   1&  0 &  0.00 &  2.04 &  3.00	&  0 &  0.00 &  0.00 &  3.00	\\
  3 &  10 &   1&  0 &  0.00 &  0.00 &  2.04	&  1 & 10.00 &  6.08 &  3.92	\\
  4 &  10 & 0.5&  1 & 10.00 &  4.41 &  5.59	&  0 &  0.00 &  0.00 &  6.08	\\
  5 &  10 & 0.5&  0 &  0.00 &  0.00 &  4.41	&  1 & 10.00 &  3.54 &  6.46	\\
  6 &  10 & 0.5&  1 &  8.22 &  0.00 &  8.22	&  0 &  0.00 &  0.00 &  3.54	\\
         \hline
    \multicolumn{3}{|c||}{~}
    & \multicolumn{4}{|c|}{$\Pi^1 = 67.13$} 
    & \multicolumn{4}{|c||}{$\Pi^2 = 65.72$} 
\\\hline  
  1 &  10 &   1&  1 &  9.00 &  5.67 &  3.33	&  1 &  9.01 &  5.67 &  3.34	\\
  2 &  10 &   1&  0 &  0.00 &  2.67 &  3.00	&  0 &  0.00 &  2.67 &  3.00	\\
  3 &  10 &   1&  0 &  0.00 &  0.00 &  2.67	&  0 &  0.00 &  0.00 &  2.67	\\
  4 &  10 & 0.5&  1 &  8.00 &  0.66 &  7.34	&  1 & 10.00 &  4.67 &  5.33	\\
  5 &  10 & 0.5&  1 & 10.00 &  4.00 &  6.67	&  0 &  0.00 &  0.00 &  4.67	\\
  6 &  10 & 0.5&  0 &  0.00 &  0.00 &  4.00	&  1 &  8.00 &  0.00 &  8.00	\\
         \hline
    \multicolumn{3}{|c||}{~}
    & \multicolumn{4}{|c|}{$\Pi^1 = 64.30$} 
    & \multicolumn{4}{|c||}{$\Pi^2 = 64.34$} 
\\\hline
  \end{tabular}
\end{scriptsize}
\caption{Different equilibria obtained for the same parameter set.}
  \label{tab:manyeq}
\end{table}

There may also be the case that the capacities are asymmetric; in this
case the firm with larger capacity has a competitive advantage, as
shown in Table~\ref{tab:asymmetric}.

\begin{table}[htbp]
  \centering
  \begin{scriptsize}
  \begin{tabular}{|c|c|c||c|r|r|r||c|r|r|r|}
    \hline
    \multicolumn{3}{|c||}{~}
    & \multicolumn{4}{|c||}{Firm 1} 
    & \multicolumn{4}{|c|}{Firm 2}\\\hline
    \hline
    \multicolumn{3}{|c||}{~}
    & \multicolumn{4}{|c||}{$F=10, H=1, K=10$} 
    & \multicolumn{4}{|c|}{$F=10, H=1, K=25$}\\\hline
Period (t) & ~$a_t$~ & ~$b_t$~
& ~~$y_t$~~ & ~~$x_t$~~ & ~~$h_t$~~ & ~~$q_t$~~
& ~~$y_t$~~ & ~~$x_t$~~ & ~~$h_t$~~ & ~~$q_t$~~ 
\\\hline  
  1 &  10 &   1&  1 &  8.33 &  5.00 &  3.33	&  1 &  6.33 &  3.00 &  3.33	\\
  2 &  10 &   1&  0 &  0.00 &  2.00 &  3.00	&  0 &  0.00 &  0.00 &  3.00	\\
  3 &  10 &   1&  0 &  0.00 &  0.00 &  2.00	&  1 &  9.34 &  5.34 &  3.99	\\
  4 &  10 & 0.5&  1 &  7.33 &  0.00 &  7.32	&  0 &  0.00 &  0.00 &  5.34	\\
  5 &  10 & 0.5&  1 & 10.00 &  4.67 &  5.34	&  1 & 13.99 &  6.67 &  7.32	\\
  6 &  10 & 0.5&  0 &  0.00 &  0.00 &  4.67	&  0 &  0.00 &  0.00 &  6.67	\\
         \hline
    \multicolumn{3}{|c||}{~}
    & \multicolumn{4}{|c|}{$\Pi^1 = 56.11$} 
    & \multicolumn{4}{|c||}{$\Pi^2 = 69.46$} 
\\\hline
\end{tabular}
\end{scriptsize}
\caption{Equilibrium with asymmetric capacities.}
  \label{tab:asymmetric}
\end{table}

Let us now turn to the case where one firm decides to play a quantity
such that the other is put out of the market, as in a Stackelberg
equilibrium.  The problem of optimally determining that (leader's)
quantity is not trivial, as a sub-problem of this is to determine the
point where profit becomes non-null in a minimum cost production plan
(this is the follower's problem).  Empirically, this equilibrium can
be determined by increasing the quantity put into the market by one
firm, until the optimal response of the opponents be to produce zero.
Results for this case are presented in Table~\ref{tab:deter}.  Note
that we are unsure if this is an equilibrium; we just verified that
any slight reduction in Firm 2's quantities induce Firm 1 to play,
resulting in a large decrease on Firm 2's profit.  In the situation
presented, Firm 1 has no incentive to play a positive amount, as it
would result in losses.

\begin{table}[htbp]
  \centering
  \begin{scriptsize}
  \begin{tabular}{|c|c|c||c|r|r|r||c|r|r|r|}
    \hline
    \multicolumn{3}{|c||}{~}
    & \multicolumn{4}{|c||}{Firm 1} 
    & \multicolumn{4}{|c|}{Firm 2}\\\hline
    \hline
    \multicolumn{3}{|c||}{~}
    & \multicolumn{4}{|c||}{$F=10, H=1, K=25$} 
    & \multicolumn{4}{|c|}{$F=10, H=1, K=25$}\\\hline
Period (t) & ~$a_t$~ & ~$b_t$~
& ~~$y_t$~~ & ~~$x_t$~~ & ~~$h_t$~~ & ~~$q_t$~~
& ~~$y_t$~~ & ~~$x_t$~~ & ~~$h_t$~~ & ~~$q_t$~~ 
\\\hline  
  1 &  10 &   1&  0 &  0.000 &  0.000 &  0.000	&  1 & 10.888 &  5.449 &  5.439	\\
  2 &  10 &   1&  0 &  0.000 &  0.000 &  0.000	&  0 &  0.000 &  0.000 &  5.449	\\
  3 &  10 &   1&  0 &  0.000 &  0.000 &  0.000	&  1 &  5.469 &  0.000 &  5.469	\\
  4 &  10 & 0.5&  0 &  0.000 &  0.000 &  0.000	&  1 & 13.409 &  0.000 & 13.409	\\
  5 &  10 & 0.5&  0 &  0.000 &  0.000 &  0.000	&  1 & 12.889 &  0.000 & 12.889	\\
  6 &  10 & 0.5&  0 &  0.000 &  0.000 &  0.000	&  1 & 12.769 &  0.000 & 12.769	\\
         \hline
    \multicolumn{3}{|c||}{~}
    & \multicolumn{4}{|c|}{$\Pi^1 = 0$} 
    & \multicolumn{4}{|c||}{$\Pi^2 = 155.119$} 
\\\hline
  \end{tabular}
\end{scriptsize}
\caption{Equilibrium when firm 2 defects (plays a quantity such that
    firm 1's optimal response is to play zero).}
  \label{tab:deter}
\end{table}

Clearly, if Firm 1 would play the same quantities as a Firm 2, it
would result in large losses for both firms.  This example serves as
an introduction to a different type of game, where firms may
collaborate (play Cournot's quantities) or defect.  This makes a
bridge between games in market equilibria and 2-person games like the
Prisoner's Dilemma.  In this context, due to the existence of several
periods, repeated games are particularly interesting; we will turn to
their analysis in the next section.

\section{Iterated games}
\label{sec:iterated}

One may consider that the decisions for all the periods have to be
taken in advance, and the corresponding quantities fixed for all the
planning horizon.  However, it is much more natural to consider that
at the begin of each period there is a commitment only regarding the
quantity to produce on that period, and that the moves concerning
later periods remain open.  This situation leads to an interesting
game between the firms, where in each period a firm may decide to
cooperate with the others, or to defect.  

In many situations, the computation of a Nash equilibrium in the
presence of discrete variables is
NP-hard~\cite{Papadimitriou2001,Daskalakis2009}.  As for our problem,
even the monopoly case is NP-hard.  The equilibria computed in the
previous section for the duopoly case are based on the assumption that
each firm knows the production decisions of the other for the whole
planning horizon.  If that information is not available (as usual in
real-world cases), the problem becomes more complicated.  As we have
seen, the solutions are many times asymmetric; there is no easy
strategy for deciding on the role of each firm if the future periods'
moves are not known.  Indeed, we are not aware of an optimal strategy
for this general case.

In order to complete the illustration, we go back to the example and
provide results for a simple strategy, on a two-firm game, equivalent
of a well known strategy in the iterated Prisoners'
Dilemma~\cite{Rapoport1965}:

\begin{enumerate}
\item On the first period cooperate: play an optimal (non-defecting)
  quantity given by equation~\ref{eq:oligopoly} if it leads to a
  positive profit, or null quantity otherwise.
\item On the subsequent periods:
  \begin{enumerate}
  \item if in the previous period the opponent cooperated, cooperate
    too: play a Cournot's quantity;
  \item otherwise, retaliate: play a large quantity, such that the
    other firms' optimal reaction would be producing zero quantity.
  \end{enumerate}
\end{enumerate}

Notice that, as in some instances there are several possibilities for
Cournot equilibria, the above strategy may be ambiguous.  This can be
observed on Table~\ref{tab:cooperate}.  Even though in the best
situation the Cournot equilibrium with future information is obtained
(top), other (inferior) situations may also arise (middle and bottom);
firms have no information to decide which plays what in asymmetric
moves.

\begin{table}[htbp]
  \centering
  \begin{scriptsize}
  \begin{tabular}{|c|c|c||c|r|r|r||c|r|r|r|}
    \hline
    \multicolumn{3}{|c||}{~}
    & \multicolumn{4}{|c||}{Firm 1} 
    & \multicolumn{4}{|c|}{Firm 2}\\\hline
    \hline
    \multicolumn{3}{|c||}{~}
    & \multicolumn{4}{|c||}{$F=10, H=1, K=25$} 
    & \multicolumn{4}{|c|}{$F=10, H=1, K=25$}\\\hline
Period (t) & ~$a_t$~ & ~$b_t$~
& ~~$y_t$~~ & ~~$x_t$~~ & ~~$h_t$~~ & ~~$q_t$~~
& ~~$y_t$~~ & ~~$x_t$~~ & ~~$h_t$~~ & ~~$q_t$~~ 
\\\hline  
  1 &  10 &   1&  1 &  8.33 &  5.00 &  3.34	&  1 &  6.33 &  3.00 &  3.34	\\
  2 &  10 &   1&  0 &  0.00 &  2.00 &  3.00	&  0 &  0.00 &  0.00 &  3.00	\\
  3 &  10 &   1&  0 &  0.00 &  0.00 &  2.00	&  1 &  9.33 &  5.33 &  4.00	\\
  4 &  10 & 0.5&  1 & 12.67 &  5.33 &  7.34	&  0 &  0.00 &  0.00 &  5.33	\\
  5 &  10 & 0.5&  0 &  0.00 &  0.00 &  5.33	&  1 & 12.66 &  5.33 &  7.33	\\
  6 &  10 & 0.5&  1 &  7.34 &  0.00 &  7.34	&  0 &  0.00 &  0.00 &  5.33	\\
         \hline
    \multicolumn{3}{|c||}{~}
    & \multicolumn{4}{|c|}{$\Pi^1 = 62.15$} 
    & \multicolumn{4}{|c||}{$\Pi^2 = 61.43$} 
\\\hline\hline
  1 &  10 &   1	&  1 &	6.33 &	3.00 &	3.34	&  1 &	6.33 &	3.00 &	3.34	\\
  2 &  10 &   1	&  0 &	0.00 &	0.00 &	3.00	&  0 &	0.00 &	0.00 &	3.00	\\
  3 &  10 &   1	&  1 &	9.33 &	5.33 &	4.00	&  1 &	9.33 &	5.33 &	4.00	\\
  4 &  10 & 0.5	&  0 &	0.00 &	0.00 &	5.33	&  0 &	0.00 &	0.00 &	5.33	\\
  5 &  10 & 0.5	&  1 & 12.66 &	5.33 &	7.33	&  1 & 12.66 &	5.33 &	7.33	\\
  6 &  10 & 0.5	&  0 &	0.00 &	0.00 &	5.33    &  0 &	0.00 &	0.00 &	5.33	\\
         \hline
    \multicolumn{3}{|c||}{~}
    & \multicolumn{4}{|c|}{$\Pi^1 = 56.70$} 
    & \multicolumn{4}{|c||}{$\Pi^2 = 56.70$} 
\\\hline\hline
  1 &  10 &   1&  1 &  8.33 &  5.00 &  3.34	&  1 &	8.33 &	5.00 &	3.34    \\
  2 &  10 &   1&  0 &  0.00 &  2.00 &  3.00	&  0 &	0.00 &	2.00 &	3.00    \\
  3 &  10 &   1&  0 &  0.00 &  0.00 &  2.00	&  0 &	0.00 &	0.00 &	2.00    \\
  4 &  10 & 0.5&  1 & 12.67 &  5.33 &  7.34	&  1 & 12.67 &	5.33 &	7.34    \\
  5 &  10 & 0.5&  0 &  0.00 &  0.00 &  5.33	&  0 &	0.00 &	0.00 &	5.33    \\
  6 &  10 & 0.5&  1 &  7.34 &  0.00 &  7.34	&  1 &	7.34 &	0.00 &	7.34    \\
         \hline
    \multicolumn{3}{|c||}{~}
    & \multicolumn{4}{|c|}{$\Pi^1 = 56.78$} 
    & \multicolumn{4}{|c||}{$\Pi^2 = 56.78$} 
\\\hline
  \end{tabular}
\end{scriptsize}
\caption{Results for repeated games: both firms cooperate.}
  \label{tab:cooperate}
\end{table}

Much inferior outcomes are obtained if one of the firms decides to
defect.  A firm might be tempted to impose a large move; but, as the
other firm has no information about this, it will play either the
``cooperate'' or the ``defect'' move, resulting in large losses for
both.  Table~\ref{tab:defect} presents these situations.  It should be
noted that, in the cases presented, the last firm defecting is worse
off at the end.  This encourages firms to defect, trying to make the
other firms  drop out of the market, and have the large profits of
Table~\ref{tab:deter}.
\begin{table}[htbp]
  \centering
  \begin{scriptsize}
  \begin{tabular}{|c|c|c||c|r|r|r||c|r|r|r|}
    \hline
    \multicolumn{3}{|c||}{~}
    & \multicolumn{4}{|c||}{Firm 1} 
    & \multicolumn{4}{|c|}{Firm 2}\\\hline
    \hline
    \multicolumn{3}{|c||}{~}
    & \multicolumn{4}{|c||}{$F=10, H=1, K=25$} 
    & \multicolumn{4}{|c|}{$F=10, H=1, K=25$}\\\hline
Period (t) & ~$a_t$~ & ~$b_t$~
& ~~$y_t$~~ & ~~$x_t$~~ & ~~$h_t$~~ & ~~$q_t$~~
& ~~$y_t$~~ & ~~$x_t$~~ & ~~$h_t$~~ & ~~$q_t$~~ 
\\\hline  
  1 &  10 &   1&  1 &   10.88 &  5.44 &	  5.44	  &  1 &   10.88 &  5.44 &  5.44 \\
  2 &  10 &   1&  1 &	 0.00 &	 0.00 &	  5.44	  &  0 &    0.00 &  0.00 &  5.44 \\
  3 &  10 &   1&  0 &	 5.47 &	 0.00 &	  5.47	  &  1 &    5.47 &  0.00 &  5.47 \\
  4 &  10 & 0.5&  1 &	13.41 &	 0.00 &	 13.41	  &  1 &   13.41 &  0.00 & 13.41 \\
  5 &  10 & 0.5&  1 &	12.89 &	 0.00 &	 12.89	  &  1 &   12.89 &  0.00 & 12.89 \\
  6 &  10 & 0.5&  1 &	12.78 &	 0.00 &	 12.78	  &  1 &   12.78 &  0.00 & 12.78 \\
         \hline
    \multicolumn{3}{|c||}{~}
    & \multicolumn{4}{|c|}{$\Pi^1 = -55.44$} 
    & \multicolumn{4}{|c||}{$\Pi^2 = -55.44$} 
\\\hline\hline
  1 &  10 &   1 &  1 &  6.33 &  3.00 &  3.34    &  1 &	 10.88 &  5.44 &  5.44	  \\
  2 &  10 &   1 &  0 &	0.00 &	0.00 &	3.00	&  0 &	  0.00 &  0.00 &  5.44	  \\
  3 &  10 &   1 &  1 &	9.33 &	5.33 &	4.00	&  1 &	  5.47 &  0.00 &  5.47	  \\
  4 &  10 & 0.5 &  0 &	0.00 &	0.00 &	5.33	&  1 &	 13.41 &  0.00 & 13.41	  \\
  5 &  10 & 0.5 &  1 & 12.66 &	5.33 &	7.33	&  1 &	 12.89 &  0.00 & 12.89	  \\
  6 &  10 & 0.5 &  0 &	0.00 &	0.00 &	5.33	&  1 &	 12.78 &  0.00 & 12.78	  \\
         \hline
    \multicolumn{3}{|c||}{~}
    & \multicolumn{4}{|c|}{$\Pi^1 = -23.77$} 
    & \multicolumn{4}{|c||}{$\Pi^2 = -15.01$} 
\\\hline\hline
  1 &  10 &   1&  1 & 8.33  & 4.99 &  3.34	&  1 &  10.88  & 5.44	&  5.44	\\
  2 &  10 &   1&  1 & 5.89  & 5.44 &  5.44	&  0 &	0.00   & 2.10	&  3.34	\\
  3 &  10 &   1&  0 & 0.00  & 2.10 &  3.34	&  1 &	3.37   & 0.00	&  5.47	\\
  4 &  10 & 0.5&  1 & 11.31 & 0.00 &  13.41	&  1 &	12.67  & 5.33	&  7.34	\\
  5 &  10 & 0.5&  1 & 12.67 & 5.33 &  7.34	&  1 &	7.56   & 0.00	&  12.89\\
  6 &  10 & 0.5&  1 & 7.45  & 0.00 &  12.78	&  1 &	7.34   & 0.00	&  7.34	\\
         \hline
    \multicolumn{3}{|c||}{~}
    & \multicolumn{4}{|c|}{$\Pi^1 = -53.17$} 
    & \multicolumn{4}{|c||}{$\Pi^2 = -45.65$} 
\\\hline
  \end{tabular}
\end{scriptsize}
\caption{Results for repeated games: both firms defect (top), Firm 2
    defects and Firm 1 cooperates (middle), 
    or play other firm's last move (bottom; Firm 2 starts defecting).}
  \label{tab:defect}
\end{table}

\section{Conclusions}
\label{sec:conclusions}

Recent advances in mixed-integer nonlinear programming software allow
the straightforward computation of optimal solution for problems which
were very difficult to solve some years ago.  This allowed us to
compute equilibria in a simple but very interesting market, where a
lot-sizing problem is setup in a competitive context.  

The analysis of a simple example puts in evidence that the solution of a
lot-sizing problem may be very different of its usual, constant demand
version.  This also extends, in a somewhat different way, recent work
on agent-based approaches for computational
equilibria~\cite{Kimbrough2008} to a recurring, dynamic environment.

When several firms play, the existence of several periods in the
lot-sizing problem makes it natural to have an iterated game where
each firm plays one move per period.  We proposed a very simple
strategy; more sophisticated ones, e.g. mixed strategies, are an
interesting subject for further research in this topic.  Another
direction concerns the analysis of real-world problems under this
background; many markets have points in common with this model (for
example, energy markets, often tackled as equilibrium
models~\cite{Ehrenmann2008}).  A scenario which is worthy studying is
the one where firms with consecutive losses drop out of the market, as
happens in most real-world cases.

The illustration presented in this paper is also interesting for
broadening the range of situations analysed in the study of equilibria
in Economics, putting in close relation economic equilibrium and
computational complexity, two subjects seldom studied together.
Iterated games are often considered excellent teaching
tools~\cite{Resler1996}; we hope that the one described here will also
make a contribution in this regard.

\paragraph*{Acknowledgment.} This research was supported in part by
FCT -- Funda{\c c}{\~a}o para a Ci{\^e}ncia e a Tecnologia (Project
**PTDC/GES/73801/2006) and by an European project under Framework
Programme 7 (CIVITAS-ELAN: Mobilising citizens for vital cities).

\bibliographystyle{splncs}
\bibliography{nash-co}

\end{document}